\documentclass[aps,prb,showpacs,reprint, superscriptaddress]{revtex4-1}
\usepackage{graphicx}
\usepackage{color}
\usepackage{amsmath}
\usepackage{bbm}
\usepackage{amssymb}
 
\usepackage[T1]{fontenc}

\input epsf

\begin{document}

\title{Stability of the coexistent superconducting-nematic phase under the presence of intersite interactions}

%\title{Effect of correlated hopping and intersite Coulomb interaction on the selected principal features of high temperature superconductors within the t-J-U model}

\author{Micha{\l} Zegrodnik}
\email{michal.zegrodnik@agh.edu.pl}
\affiliation{Academic Centre for Materials and Nanotechnology, AGH University of Science and Technology, Al. Mickiewicza 30, 30-059 Krak\'ow,
Poland}

\author{J{\'o}zef Spa{\l}ek}
\email{jozef.spalek@uj.edu.pl}
\affiliation{Marian Smoluchowski Institute of Physics, 
Jagiellonian University, ul. \L ojasiewicza 11,
30-348 Krak\'ow, Poland}

\begin{abstract}
We analyze the effect of intersite-interaction terms on the stability of the coexisting superconucting-nematic phase (SC+N) within the extended Hubbard and $t$-$J$-$U$ models on the square lattice. In order to take into account the correlation effects with a proper precision, we use the approach based on the \textit{diagrammatic expansion of the Gutzwiller wave function} (DE-GWF), which goes beyond the renormalized mean field theory (RMFT) in a systematic manner. As a starting point of our analysis we discuss the stability region of the SC+N phase on the intrasite Coulomb repulsion-hole doping plane for the case of the Hubbard model. Next, we show that the  exchange interaction term enhances superconductivity while suppresses the nematicity, whereas the intersite Coulomb repulsion term acts in the opposite manner. The competing character of the SC and N phases interplay is clearly visible throughout the analysis. A universal conclusion is that the nematic phase does not survive within the $t$-$J$-$U$ model 
with the value of $J$ integral typical for the high-T$_C$ cuprates ($J\approx 0.1$eV). For the sake of completeness, the effect of the 
correlated hopping term is also 
analyzed. Thus the present discussion contains all relevant two-site interaction terms which appear in the parametrized one-band model within the second quantization scheme. At the end, the influence of the higher-order terms of the diagrammatic expansion on the rotational symmetry breaking is also shown by comparing the DE-GWF results with those corresponding to the RMFT.

\end{abstract}

\maketitle

\section{Introduction}\label{sec:intro}
The nematic ordering is believed to appear in a number of strongly correlated compounds such as URu$_2$Si$_2$\cite{Okazaki2011}, iron-pnictides\cite{Chu2012,Fernandes2014}, cuprates\cite{Keimer2015,Pelc2016,Lawler2010}, Sr$_3$Ru$_2$O$_7$\cite{Borzi2007}, as well as quantum Hall systems\cite{Lilly1999}. Nematicity is characterized by a spontaneous rotational symmetry breaking of the electronic structure, with the preservation of the translational symmetry imposed by the crystal lattice. This condition excludes positional or magnetic orderings such as those appearing in the cases of spin-density-wave (SDW) or charge-density-wave (CDW) phases. However, it has been argued that in the cuprates the CDW phase may be formed through a precursor state which has a nematic character\cite{Pelc2016}. In some of the copper based compounds a small anisotropy of the lattice makes it difficult to validate the nematic behavior of the electronic wave function, as the C$_4$ symmetry of the Cu-O planes is already broken by the 
crystal structure. Nevertheless, in spite of such a small structural 
anisotropy, a large anisotropy 
of various physical properties has been observed\cite{Ando2002,Daou2010,Hinkov2008,Lawler2010}. This fact, together with recent research on  LESCO, LNSCO, and LBCO compounds\cite{Pelc2016,Achkar2016} indicate, that the anisotropic character of electronic properties of Cu-O planes is not a trivial consequence of the lattice distortions. Instead, it may be caused by an intrinsic susceptibility towards the nematic order appearance and may be due to the inter-electronic interactions.

For the copper-based materials the appearance of superconducting phase can also be ascribed to the interelectronic correlation effects. Therefore, the question of the SC and N phases interplay/coexistence within typical models referring to strongly correlated systems is worth exploring. The mean-field analysis of SC+N appearance for the case of phenomenological model suggests that the two phases compete with each other\cite{Yamase2007}. Other investigations, going beyond the mean-field approach, included methods limited only to weak or intermediate interactions\cite{Neumayr2003,Kitatani2017,Honerckamp2002,Hankevych2002}. The SC+N phase induced solely by strong correlations has been analyzed recently\cite{Kaczmarczyk2016} for the case of Hubbard model (with intrasite repulsion only), by using the \textit{diagrammatic expansion of the Guwtziller wave function} (DE-GWF) approach. The same method has been applied by us to the $t$-$J$-$U$ 
model what has lead to a very good quantitative agreement between theory and 
experiment for the selected principal properties of the superconducting phase in the cuprates\cite{Spalek2017,Zegrodnik2017_1}. Namely, it has been found that, the presence of both the $J$ term and the possibility of having a small but non-zero number of double occupancies at the same time  was indispensable in order to obtain the proper quantitative agreement. One should also note that additional interactions terms which are frequently omitted, may affect the stability of various correlation-induced phases\cite{Zegrodnik2017_2,Abram2017}.

Here we use the DE-GWF method in order to carry out a detailed analysis of nematic and superconducting phases coexistence/competition in the presence of all significant two-site interaction terms, i.e., the antiferromagnetic exchange, the intersite Coulomb repulsion, and the correlated hopping. To show that the C$_4$ symmetry breaking presented here is due to interelectronic effects, we focus mainly on the square-lattice structure. However, the influence of the preexistent lattice distortion is also discussed. To show that the higher order terms of the diagrammatic expansion are essential to induce the tendency towards the spontaneous C$_4$ symmetry breaking, we compare the obtained results with those calculated within the RMFT method equivalent to the zeroth order expansion of the GWF\cite{Kaczmarczyk2014}. 

The structure of the paper is as follows. In the next Section we introduce the $t$-$J$-$U$-$V$ model and the DE-GWF method of its solution. In Sec. III we discuss the resulting phase diagram and related physical properties comprising the regimes of pure- and coexisting-phases stability. Conclusions are contained in Sec. IV.

\section{Model and Method}\label{sec:theory}

The most general form of the Hamiltonian considered here is given below
\begin{equation}
\begin{split}
\mathcal{\hat{H}}&=\sum_{\langle ij\rangle\sigma}\big[t+K(\hat{n}_{i\bar{\sigma}}+\hat{n}_{j\bar{\sigma}})\big]\hat{c}^{\dagger}_{i\sigma}\hat{c}_{j\sigma}
+t'\sum_{\langle\langle ij\rangle\rangle\sigma}\hat{c}^{\dagger}_{i\sigma}\hat{c}_{j\sigma}\\
&+J\sum_{\langle ij\rangle}\hat{\mathbf{S}}_i\cdot\hat{\mathbf{S}}_j
+U\sum_i \hat{n}_{i\uparrow}\hat{n}_{i\downarrow}+V\sum_{\langle ij\rangle} \hat{n}_{i}\hat{n}_{j}.
 \label{eq:H_start}
 \end{split}
\end{equation}
The first two terms contain the single-particle and the correlated-hopping ($\sim K$) contributions, respectively, the third term represents the antiferromagnetic exchange interaction, and the last two terms refer to the intra- and inter-site Coulomb repulsions, respectively. By $\langle...\rangle$ and $\langle\langle...\rangle\rangle$ we denote the summations over the nearest-neighbors and next-nearest-neighbors, respectively. For $J=K=V\equiv0$ we obtain the Hubbard model which constitutes the reference point of our analysis of the particular interaction terms and their influence on the SC+N phase. With the increasing $U\rightarrow \infty$ the model reduces to an extended $t$-$J$ model.

In order to take into account the inter-electronic correlations we use the description based on the Gutzwiller-type wave function defined by
\begin{equation}
 |\Psi_G\rangle\equiv\hat{P}_G|\Psi_0\rangle,
\end{equation}
where $|\Psi_0\rangle$ is the non-correlated wave function (to be defined later) and the correlation operator $\hat{P}_G$ is provided below 
 \begin{equation}
 \hat{P}_G\equiv\prod_i\hat{P}_i=\prod_i\sum_{\Gamma}\lambda_{i,\Gamma} |\Gamma\rangle_{ii}\langle\Gamma|,
  \label{eq:Gutz_operator}
\end{equation}
where $\lambda_{i,\Gamma}\in\{\lambda_{i\emptyset},\lambda_{i\uparrow},\lambda_{i\downarrow},\lambda_{i d}\}$ are the variational parameters which correspond to four states of the local basis $|\emptyset\rangle_i\;, |\uparrow\rangle_i\;, |\downarrow\rangle_i\;, |\uparrow\downarrow\rangle_i$ at site $i$, respectively. An important step of the DE-GWF method is the application of the condition \cite{Bunemann2012}
\begin{equation}
 \hat{P}_i^2\equiv1+x\hat{d}^{\textrm{HF}}_i,
 \label{eq:constraint}
 \end{equation}
 where $x$ is yet another variational parameter and $\hat{d}^{\textrm{HF}}_i\equiv\hat{n}_{i\uparrow}^{\textrm{HF}}\hat{n}_{i\downarrow}^{\textrm{HF}}$, $\hat{n}_{i\sigma}^{\textrm{HF}}\equiv\hat{n}_{i\sigma}-n_{0}$, with $n_{0}\equiv\langle\Psi_0|\hat{n}_{i\sigma}|\Psi_0\rangle$. One should note that $\lambda_{\Gamma}$ parameters are all functions of $x$ which results in only one variational parameter of the wave function. As it has been shown in Refs. \onlinecite{Bunemann2012,Gebhard1990}, condition (\ref{eq:constraint}) leads to rapid convergence of the resulting diagrammatic expansion with the increasing order in the resultant variational parameter $x$.

 Within this approach, the expectation value in the correlated state from any two local operators, $\hat{o}_i$ and $\hat{o}^{\prime}_j$, can be expressed in the following form
 \begin{equation}
  \langle\Psi_G|\hat{o}_{i}\hat{o}^{\prime}_{j}|\Psi_G\rangle=\sum_{k=0}^{\infty}\frac{x^k}{k!}\sideset{}{'}\sum_{l_1...l_k}\langle\Psi_0| \tilde{o}_{i}\tilde{o}^{\prime}_{j}\;\hat{d}^{\textrm{HF}}_{l_1...l_k}|\Psi_0 \rangle,
\label{eq:expansion}
\end{equation}
where $\tilde{o}_{i}\equiv\hat{P}_i\hat{o}_{i}\hat{P}_{i}$, $\tilde{o}^{\prime}_{j}\equiv\hat{P}_j\hat{o}^{\prime}_{j}\hat{P}_{j}$, $\hat{d}^{\textrm{HF}}_{l_1...l_k}\equiv\hat{d}^{\textrm{HF}}_{l_1}...\hat{d}^{\textrm{HF}}_{l_k}$, with  $\hat{d}^{\textrm{HF}}_{\varnothing}\equiv 1$. The primmed summation has the restrictions $l_p\neq l_{p'}$, $l_p\neq i,j$ for all $p$ and $p'$. 

The averages in the non-correlated state on the right-hand side of Eq. (\ref{eq:expansion}) can be decomposed by the use of the Wick's theorem applied directly in real space and expressed in terms of the correlation functions $P_{ij} \equiv \langle \hat{c}^{\dagger}_{i\sigma} \hat{c}_{j\sigma}\rangle_0$ and $S_{ij} \equiv \langle \hat{c}^{\dagger}_{i\uparrow} \hat{c}^{\dagger}_{j\downarrow}\rangle_0$. Such a procedure allows us to express the ground state energy $\langle\mathcal{\hat{H}}\rangle_G\equiv\langle\Psi_G|\mathcal{\hat H}|\Psi_G\rangle/\langle\Psi_G|\Psi_G\rangle$ as a function of $P_{ij}$, $S_{ij}$, $n_{0}$, and $x$. It has been shown that the desirable convergence can be achieved by taking the first 4-6 terms of the expansion in 
$x$ appearing in Eq. (\ref{eq:expansion}).  Here the first $5$ terms of the diagrammatic expansion (\ref{eq:expansion}) have been taken into account when carrying out the calculations.

The effective Schr\"odinger equation can be derived from the minimization condition of the ground-state energy functional $\mathcal{F}\equiv\langle\mathcal{\hat{H}}\rangle_G-\mu_G\langle\hat{N}\rangle_G$, where $\mu_G$ and $\langle \hat{N}\rangle_G$ are the chemical potential and the total number of particles determined in the state $|\Psi_G\rangle$, respectively\cite{Schickling2014,Seibold2008}. The explicit form the equation is given below
\begin{equation}
 \hat{\mathcal{H}}_{\textrm{eff}}|\Psi_0\rangle=E|\Psi_0\rangle,
\end{equation}
where the effective single-particle Hamiltonian has the form
\begin{equation}
 \hat{\mathcal{H}}_{\textrm{eff}}=\sum_{ij\sigma}t^{\textrm{eff}}_{ij}\hat{c}^{\dagger}_{i\sigma}\hat{c}_{j\sigma}+\sum_{ij}\big(\Delta^{\textrm{eff}}_{ij}\hat{c}^{\dagger}_{i\uparrow}\hat{c}^{\dagger}_{j\downarrow}+H.c.\big),
 \label{eq:H_effective}
\end{equation}
with the effective parameters 
\begin{equation}
 t^{\textrm{eff}}_{ij}\equiv \frac{\partial\mathcal{F}}{\partial P_{ij}},\quad \Delta^{\textrm{eff}}_{ij}\equiv \frac{\partial\mathcal{F}}{\partial S_{ij}}.
 \label{eq:effective_param}
\end{equation}

It is necessary to introduce the real-space cutoff for the parameters $P_{ij}$ and $S_{ij}$, which are going to be taken into account while executing explicitly the Wick's decomposition of expansion (\ref{eq:expansion}). Here, in order to carry out calculations in a reasonable time the maximum distance has been taken as $R^2_{\textrm{max}}=5a^2$, where $a$ is the lattice constant.

The self consistent equations for all the parameters $S_{ij}$ and $P_{ij}$ are derived after transforming the effective Hamiltonian (\ref{eq:H_effective}) to the reciprocal space. The solution of self-consistent equations is concomitant with the minimization over variational parameter $x$. After calculating $P_{ij}$, $S_{ij}$, $x$, $\mu_G$, and $P_{ii}=n_0$ for a selected set of microscopic parameters ($t'$, $K$, $J$, $U$, $V$), we can determine the value of the so-called correlated SC gaps $\Delta_{G,ij}\equiv\langle \hat{c}^{\dagger}_{i\uparrow}\hat{c}^{\dagger}_{j\downarrow}\rangle_G$, as well as the correlated-hopping averages $P_{G,ij}\equiv\langle \hat{c}^{\dagger}_{i\sigma}\hat{c}_{j\sigma}\rangle_G$.

The $d$-$wave$ gap symmetry is most widely used for the description of high-T$_C$ superconductivity in the cuprates. Here, small corrections to the bare $d$-$wave$ symmetry appear due to the fact that farther-distance than the nearest averages are included, i.e., those corresponding to atomic sites up to $|\mathbf{R}_{ij}|^2\equiv|\mathbf{R}_i-\mathbf{R}_j|^2=5a^2$. In spite of that, the dominant contribution to the pairing amplitude arises from the nearest-neighbor SC averages: $\Delta^G_{1,0}$, $\Delta^G_{-1,0}$, $\Delta^G_{0,1}$, $\Delta^G_{0,-1}$, where $\Delta^G_{X,Y}\equiv \langle\hat{c}^{\dagger}_{i\uparrow}\hat{c}^{\dagger}_{j\downarrow}\rangle_G$ for $\mathbf{R}_{ij}=(X,Y)a$. For the bare $d$-$wave$ symmetry, the following conditions are fulfilled $\Delta^G_{1,0}=\Delta^G_{-1,0}$, $\Delta^G_{0,1}=\Delta^G_{0,-1}$, and $\Delta^G_{1,0}=-\Delta^G_{0,1}$. However, in general, when the C$_4$ symmetry is broken, an $s$-$wave$ admixture to the $d$-$wave$ component appears. In such a situation it is 
convenient to introduce the $d$-
$wave$ and $s$-$wave$ correlated gap parameters, respectively
\begin{equation}
\begin{split}
 \Delta^G_d&=\frac{1}{2}(\Delta^G_{1,0}-\Delta^G_{0,1}),\\
 \Delta^G_s&=\frac{1}{2}(\Delta^G_{1,0}+\Delta^G_{0,1}).
 \end{split}
\end{equation}
Also, since for the nematic phase the $(1,0)$ and $(0,1)$ directions are not equivalent, the corresponding hopping averages will also differ and the following parameter charactering the nematicity can be introduced in the form: $\delta P_G\equiv P^G_{1,0}-P^G_{0,1}$, where $P^G_{X,Y}\equiv\langle\hat{c}^{\dagger}_{i\sigma}\hat{c}_{j\sigma}\rangle_G$, for $\mathbf{R}_{ij}=(X,Y)a$. 

In the pure SC phase $\Delta^G_d\neq 0$, $\Delta^G_s\equiv 0$, and $\delta P_G\equiv 0$, whereas in the coexistent SC+N phase: $\Delta^G_d\neq 0$, $\Delta^G_s\neq 0$, and $\delta P_G\neq 0$. For the case of pure nematic phase (without the SC order) one obtains $\Delta^G_d=\Delta^G_s\equiv0$ and $\delta P^G\neq 0$, while for the pure paramagnetic (normal) phase with neither SC nor N we have that $\Delta^G_d=\Delta^G_s\equiv0$ and $\delta P_G\equiv0$. In what follows we study systematically the phase diagram involving all the mentioned phases.

\section{Results}
In our analysis we have selected the hopping parameters as $t=-0.35$eV and $t'=0.25|t|$ (unless stated otherwise) which are typical for the copper based compounds. All the energies in the presented results are in the units of nearest-neighbor hopping integral $|t|$. The calculations correspond to the case of square lattice. However, at the end we also discuss the influence of the lattice distortion towards the orthorhombic structure.

\begin{figure}[h!]
\centering
 \includegraphics[width=0.45\textwidth]{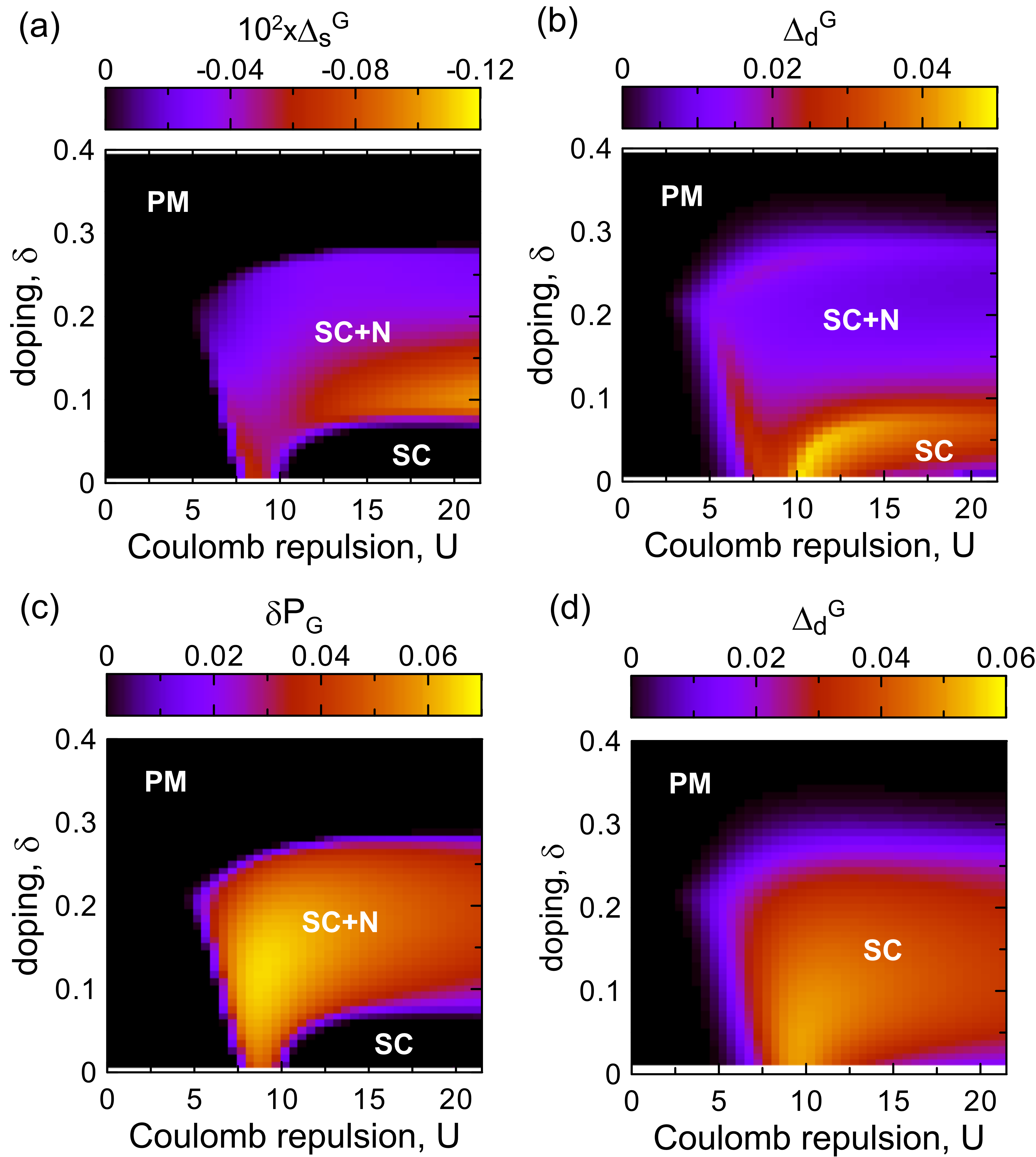}
\caption{(Colors online) $s$-$wave$ (a) and $d$-$wave$ (b) components of the correlated gap, as well as the nematicity parameter $\delta P_G=P^G_{1,0}-P^G_{0,1}$ (c), all as functions of hole doping $\delta$ and the intrasite Coulomb repulsion $U$. The region with nonzero $\Delta^G_{s}$ and $\delta P_G$ corresponds to the coexistent SC+N phase, whereas the pure SC phase is characterized by $\Delta^G_s=0$, $\delta P_G=0$ and $\Delta^G_d\neq 0$. For the paramagnetic phase (PM) $\Delta^G_s=\delta P_G=\Delta^G_d=0$. For comparison, in (d) we show the $d$-$wave$ wave correlated gap for the case when the nematic phase was not included in the calculations. The results are for the Hubbard model with $J=K=V=0$.}
\label{fig:diag_nU}
\end{figure}

First, we analyze the SC+N phase coexistence in the Hubbard model defined by Hamiltonian (\ref{eq:H_start}), i.e., with $J=K=V=0$ and for the case of square lattice. These results constitute the reference point for the subsequent analysis focused on the influence of particular two-site terms on the onset of nematicity in the extended models. In Fig. \ref{fig:diag_nU} we plot the phase diagrams on the $(U,\delta)$ plane, in which we mark the stability region of the nematic phase coexistent with superconductivity (region labelled by SC+N, with $\Delta^G_s\neq 0$, $\delta P_G\neq 0$, and $\Delta^G_d\neq 0$). As one can see, the appearance of the $(1,0)$ and $(0,1)$ directions inequivalence, which manifests itself by the nonzero values of $\delta P_G$ [shown in Fig. \ref{fig:diag_nU} (c)], is accompanied by comcomitant appearance of the $s$-$wave$ component of the SC correlated gap [shown in Fig. \ref{fig:diag_nU} (a)]. However, the $s$-$wave$ gap amplitude is two orders of magnitude smaller than that 
corresponding to the $d$-$wave$ symmetry. For large values of Coulomb repulsion ($U\gtrsim 10$) superconductivity wins with the nematic 
phase in the underdoped regime and appears in the pure $d$-$wave$ form [region labelled by SC in Fig. \ref{fig:diag_nU} (a), (b), and (c)]. 
 %%%%%%%%%%%%%%%%%%%%%%%%%%%%%%%%FIG1%%%%%%%%%%%%%%%%%%%%%%%%%%%%%%%%%%%%%%%%%%%
%%%%%%%%%%%%%%%%%%%%%%%%%%%%%%%%%%%%%%%%%%%%%%%%%%%%%%%%%%%%%%%%%%%%%%%%%%%%%%%
For comparison, in Fig. \ref{fig:diag_nU} (d) we show the correlated gap for the case when the nematic phase is not taken into account. In such a situation only $d$-$wave$ component of the SC gap appears and its values are significantly larger as compared to that in the SC+N phase [c.f. Figs. \ref{fig:diag_nU} (b) and (d)]. This means that the adjustment of the SC phase to the $C_4$ symmetry breaking leads to the weakening of the $d$-$wave$ SC, what in turn indicates the competing character of SC and N phases interplay.

\begin{figure}[h!]
\centering
\includegraphics[width=0.45\textwidth]{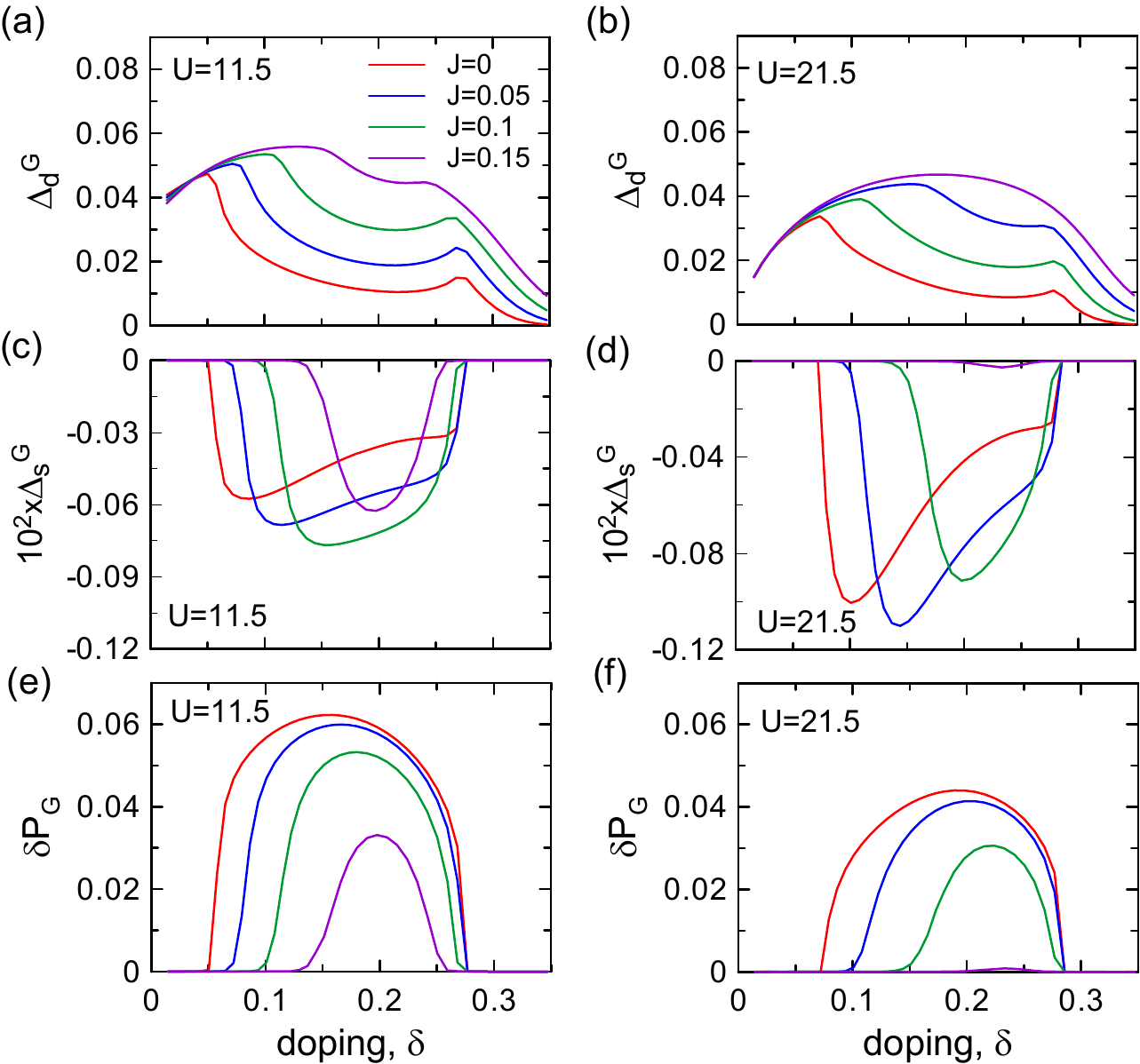}
\caption{(Colors online) $d$-$wave$ (a), (b) and $s$-$wave$ (c), (d) components of the correlated gap, as well as the nematicity parameter $\delta P_G=P^G_{1,0}-P^G_{0,1}$ (e), (f), all versus hole doping for selected values of the $J$ and $U$ parameters. One should note that with increasing $J$ the nematicity is destroyed. The results are for $V=K=0$.}
\label{fig:J_dep}
\end{figure}

In Fig. \ref{fig:J_dep} we analyze the effect of the $J$-term on the $C_4$ symmetry breaking for two significantly different values of Hubbard $U$ ($U=11.5$ and $U=21.5$). As one can see, with increasing $J$ the $d$-$wave$ superconductivity is enhanced while the nematicity gets reduced substantially. Above the value of $J\approx 0.15$ the latter is completely destroyed leaving only the pure SC phase without any $s$-$wave$ component of the gap. For larger $U$ values [Figs. (b), (d), (f)] the effect of N phase suppression is even stronger. As a result, the nematicity is already destroyed for the set of parameters for which the proper agreement between theory and experiment has been obtained in Ref. \onlinecite{Spalek2017} with respect to high-T$_C$ superconductivity in the copper-based compounds ($t=-0.35$eV, $t'=0.25|t|$, $U=22$, $J=0.25|t|$). Hence, the latter results are not affected by the $s$-$wave$ SC gap component appearance which would be destructive for the nodal direction presence observed in the 
cuprates.
%%%%%%%%%%%%%%%%%%%%%%%%%%%%%%%%FIG2%%%%%%%%%%%%%%%%%%%%%%%%%%%%%%%%%%%%%%%%%%%
%%%%%%%%%%%%%%%%%%%%%%%%%%%%%%%%%%%%%%%%%%%%%%%%%%%%%%%%%%%%%%%%%%%%%%%%%%%%%%%

\begin{figure}[h!]
\centering
\includegraphics[width=0.45\textwidth]{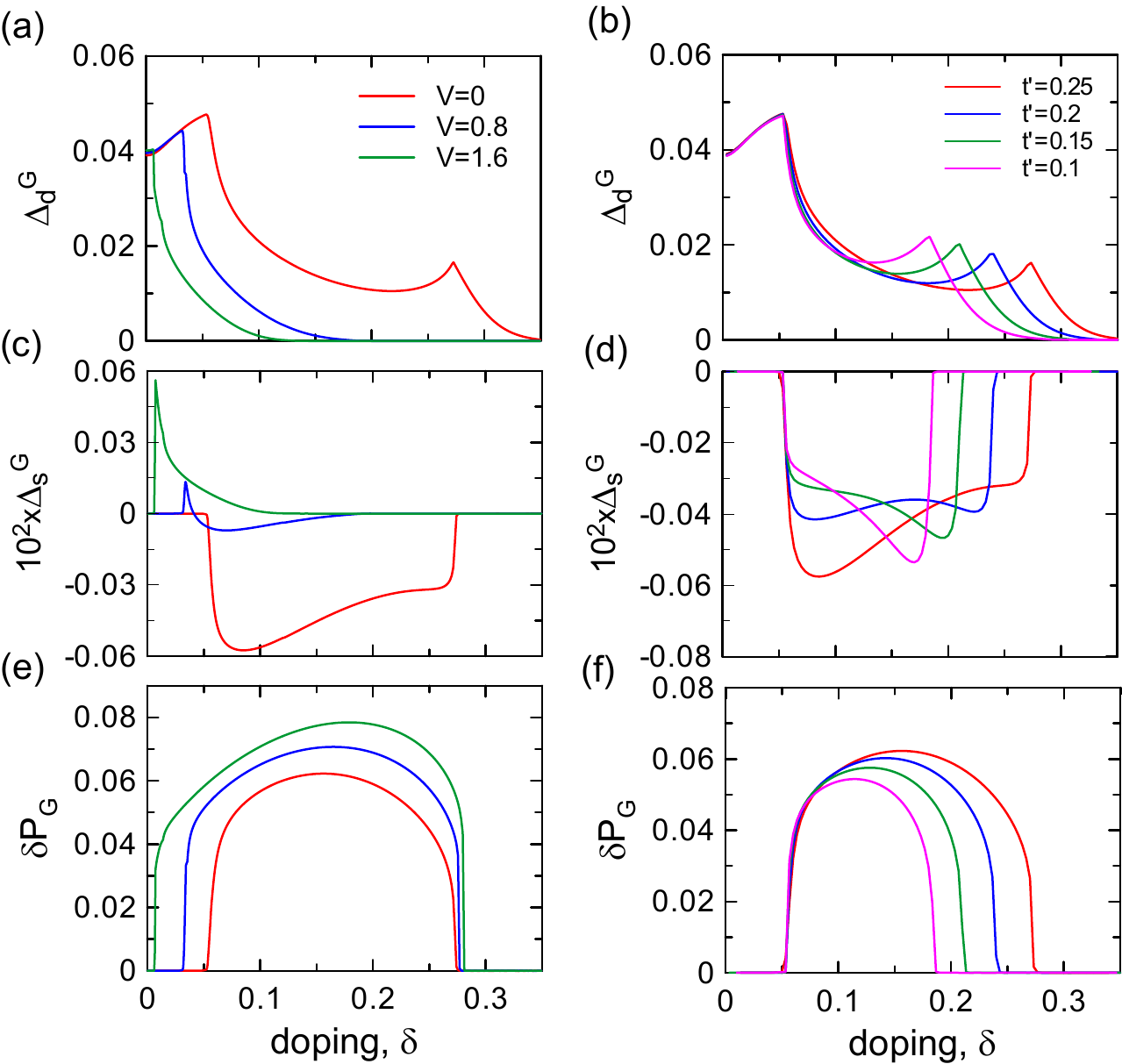}
\caption{(Colors online) $d$-$wave$ (a), (b) and $s$-$wave$ (c), (d) components of the correlated gap, as well as the nematicity parameter $\delta P_G=P^G_{1,0}-P^G_{0,1}$ (e), (f), all as functions of hole doping for selected values of $V$ and $t'$ and for $J=K=0$, $U=11.5$. One should note that with increasing $V$ the nematicity is enhanced [see (e)], whereas the SC phase is suppressed [see (a)]. Figures (a), (c), (e) correspond to $t'=0.25$, while Figs. (b), (d), (f) correspond to the $V=0$ case.}
\label{fig:V_dep}
\end{figure}

\begin{figure}[h!]
\centering
\includegraphics[width=0.5\textwidth]{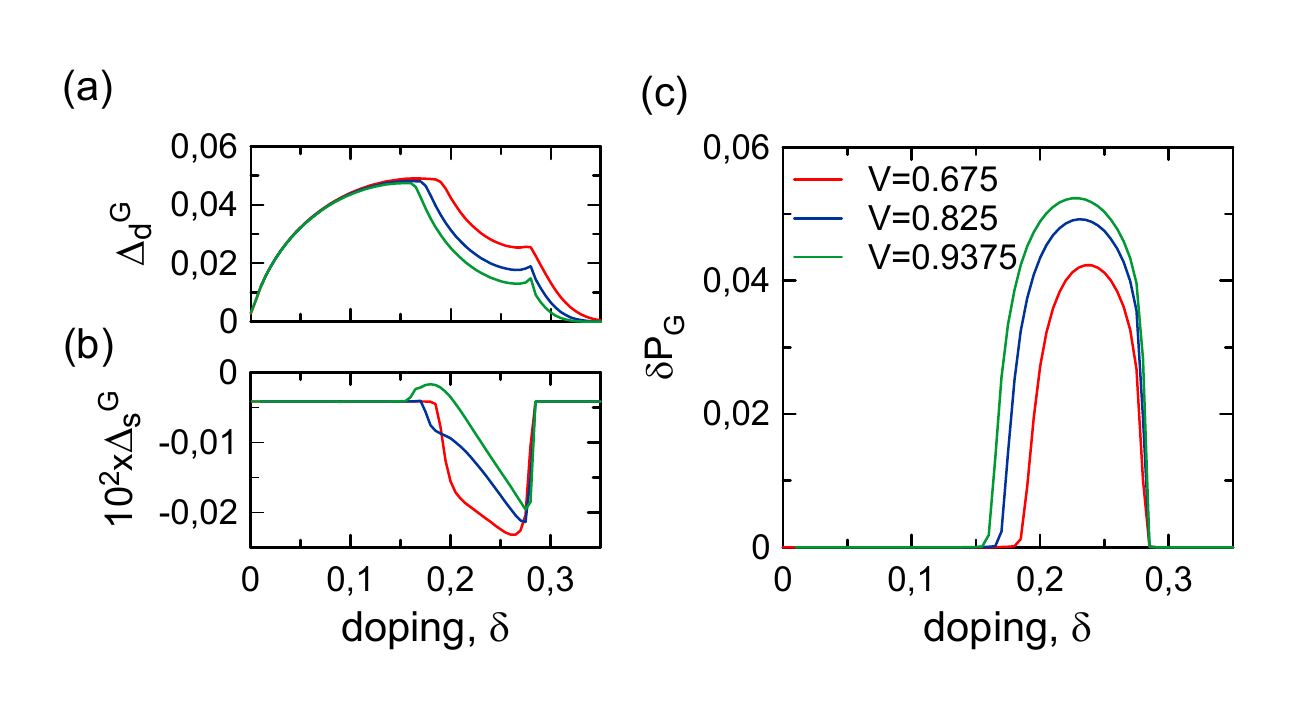}
\caption{(Colors online) $d$-$wave$ (a) and $s$-$wave$ (b) components of the correlated gap, as well as the nematicity parameter $\delta P_G=P^G_{1,0}-P^G_{0,1}$ (c) all as functions of hole doping for selected values of intersite Coulomb repulsion integral $V$ and for $U=20$, $J=0.3$.}
\label{fig:tJUV}
\end{figure}

The intersite-Coulomb repulsion term acts in the opposite manner than the $J$-term. Namely, it suppresses the pairing [see Figs. \ref{fig:V_dep} (a) and (c)] while enhances the nematicity [see Fig.  \ref{fig:V_dep} (e)]. Therefore, in the model with both $J$- and $V$-terms included, the competition between N and SC phases is determined by both these factors. As a consequence, the SC+N phase can be sustained for values of $J$ typical for the cuprates ($J\approx 0.3$) when sufficiently strong intersite Coulomb integral is considered. In Fig. \ref{fig:tJUV} we show such a situation which represent the $t$-$J$-$U$-$V$ model case. However, here the nematicity appears in the overdoped regime which would be against the experimental findings for the cuprates.    

For the sake of completeness we also analyze the influence of electronic-structure details on the SC+N phase stability. Namely, in Figs. \ref{fig:V_dep} (b), (d), and (f) we show the doping dependences of the correlated gap components and the nematicity factor, all as functions of hole doping for selected values of the next-nearest-neighbor hopping integral $t'$. As one can see, with the decreasing $t'$ value the stability region of the SC+N phase is narrowed down. However, the lower critical concentration of the nematicity onset is not affected and is close to $\delta=0.05$ [see Fig. \ref{fig:V_dep} (f)], which is similar to the upper critical concentration for the AF phase appearance observed in experiments on the cuprates. Such result differs from the one obtained recently in Ref. \onlinecite{Kitatani2017}, where it was shown that the lower critical concentration for the N phase appearance is moving together with the filling value (tuned by $t'$) which corresponds to the van Hove singularity. 
This discrepancy can be 
caused by the differences of the details of the two approaches. Namely, in the above mentioned work the FLEX+DMFT method has been used in the intermediate correlations regime ($U=4$) and at higher temperature $\beta t=20$.

%%%%%%%%%%%%%%%%%%%%%%%%%%%%%%%%FIG3%%%%%%%%%%%%%%%%%%%%%%%%%%%%%%%%%%%%%%%%%%%
\begin{figure}[h!]
\centering
\includegraphics[width=0.45\textwidth]{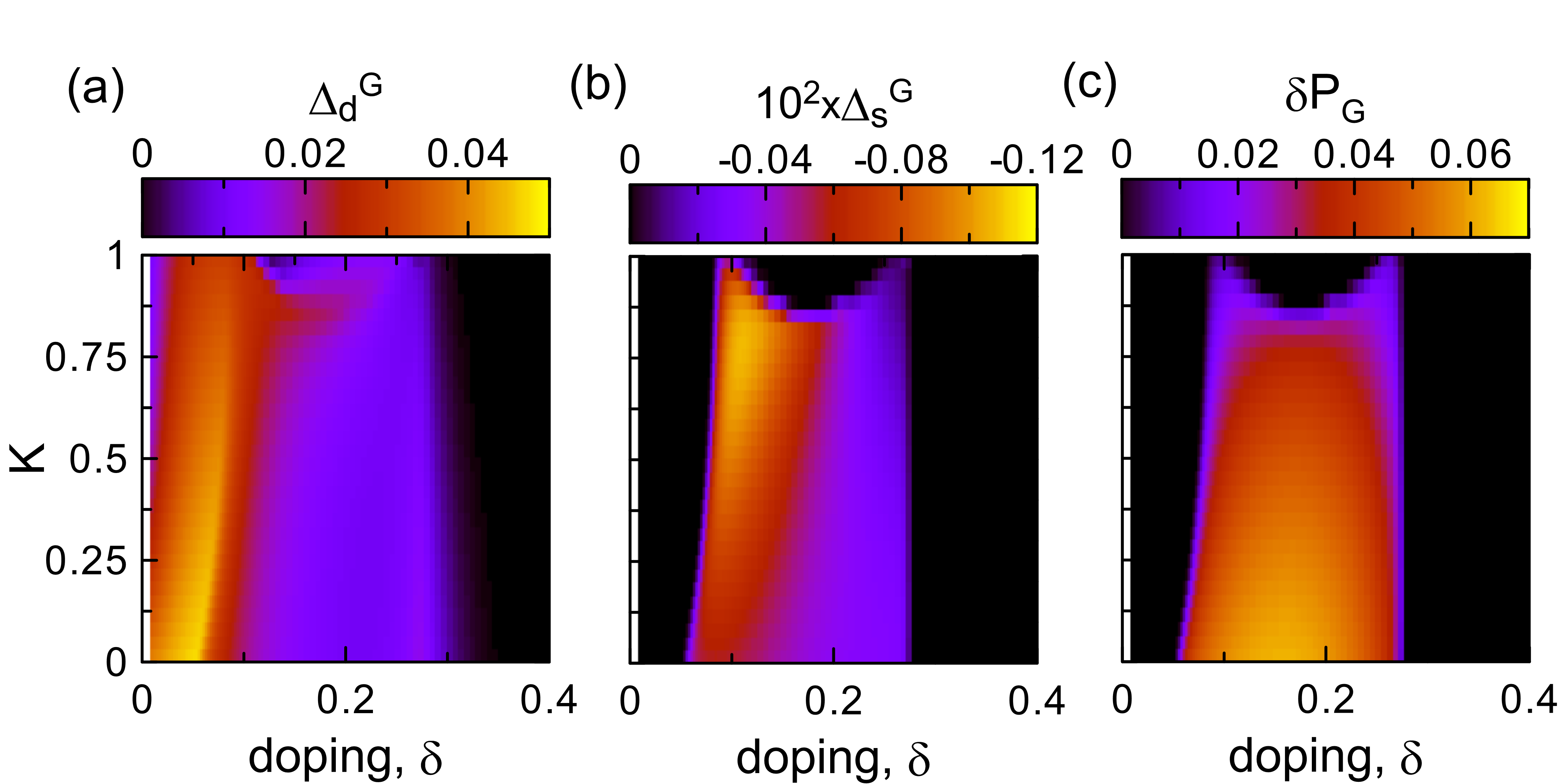}
\caption{(Colors online) $d$-$wave$ (a), and $s$-$wave$ (b) components of the correlated gap, as well as the nematicity parameter (c), all as functions of hole doping and correlated hopping integral $K$, for $U=11.5$ and $J=V=0$.}
\label{fig:K_tp_dep}
\end{figure}

%%%%%%%%%%%%%%%%%%%%%%%%%%%%%%%%%%%%%%%%%%%%%%%%%%%%%%%%%%%%%%%%%%%%%%%%%%%%%%%
The off-diagonal elements of the Coulomb interaction between the nearest neighboring lattice sites $\langle i,j\rangle$, with the corresponding two-site integral $K_{ij}\equiv\langle\mathbf{i}\mathbf{i}|V(\mathbf{r}-\mathbf{r}')|\mathbf{i}\mathbf{j}\rangle$ introduce the so-called correlated hopping term which also has been studied by us\cite{Zegrodnik2017_2}. In Fig. \ref{fig:K_tp_dep} we show the parameters which characterize the SC+N phase as functions of both hole doping $\delta$ and the correlated hopping integral $K$. As one can see the influence is not significant up to the values of $K\approx 1$, close to which the nematicity is destroyed and the SC order is being reduced.

%%%%%%%%%%%%%%%%%%%%%%%%%%%%%%%%FIG3%%%%%%%%%%%%%%%%%%%%%%%%%%%%%%%%%%%%%%%%%%%
%%%%%%%%%%%%%%%%%%%%%%%%%%%%%%%%%%%%%%%%%%%%%%%%%%%%%%%%%%%%%%%%%%%%%%%%%%%%%%%

\begin{figure}[h!]
\centering
\includegraphics[width=0.45\textwidth]{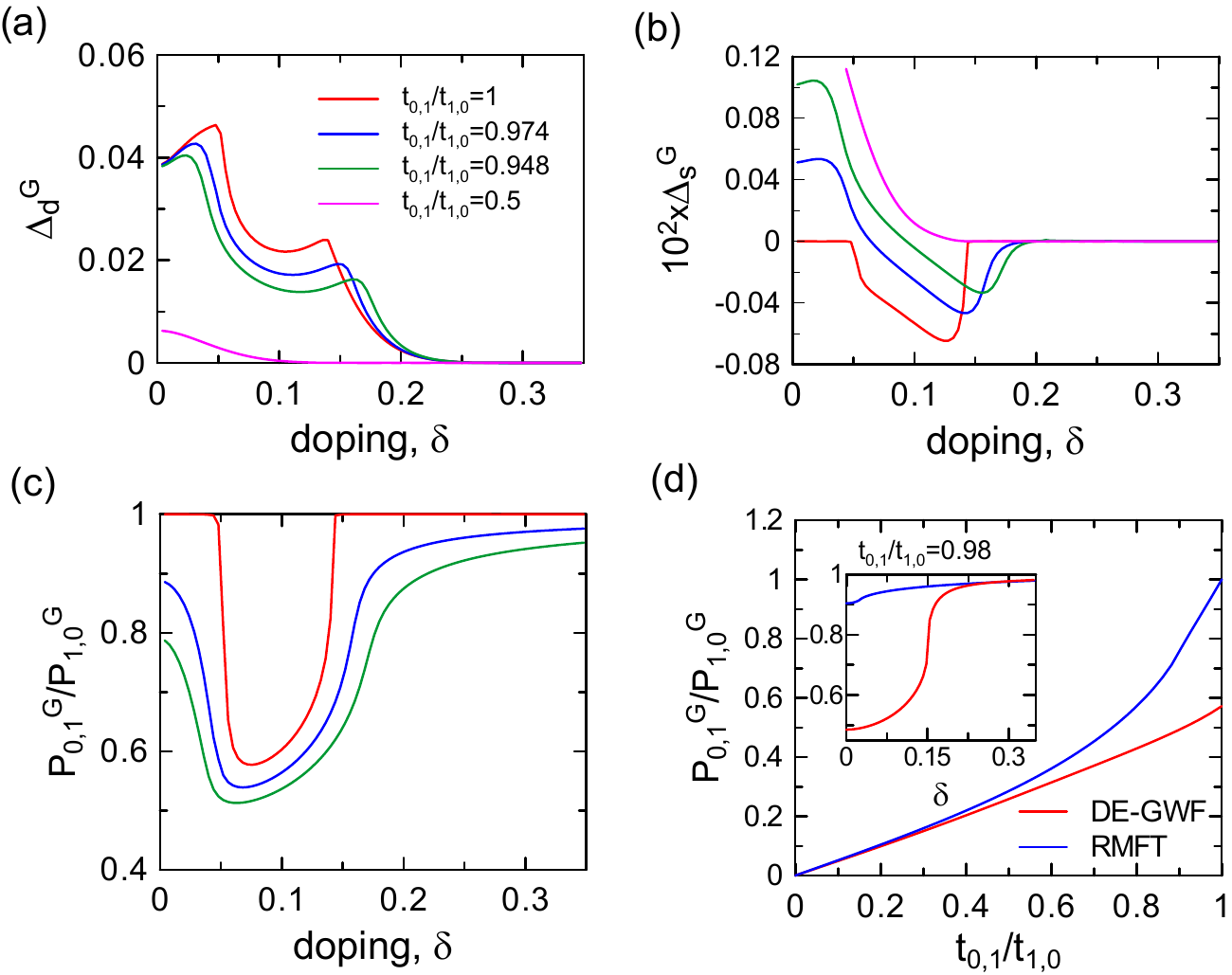}

%\epsfxsize=87mm 
%\epsfbox[147 402 540 688]{fig_dt_eps.eps}
\caption{(Colors online) $d$-$wave$ (a), and $s$-$wave$ (b), components of the correlated gap, as well as $P^G_{0,1}/P^G_{1,0}$ (c) all as functions of hole doping for  different values of the lattice distortion rate, $t_{0,1}/t_{0,1}$. For $t_{0,1}/t_{1,0}<1$ the lattice distortion is introduced which enhances nematicity and suppresses $d$-$wave$ superconductivity. In (d) we show $P^G_{0,1}/P^G_{1,0}$ for the case of pure nematic phase for $\delta=0.1$ as a function of the lattice distortion rate for the case of DE-GWF and RMFT calculations. The inset shows the doping dependence of $P^G_{0,1}/P^G_{1,0}$ for the case of pure nematic phase for the selected value of $t_{0,1}/t_{1,0}=0.98$. The results are for the Hubbard model ($J=V=K=0$) with $U=11.5$.}
\label{fig:dt_dep}
\end{figure}

It should be noted that within the present approach the appearance of the nematic phase is not induced by any straightforward mechanism such as the lattice distortion. Instead, the $C_4$ symmetry of the electronic wave function is broken spontaneously for high enough values of the Hubbard $U$. Nevertheless, in Fig. \ref{fig:dt_dep} we also provide the results with inclusion of the orthorhombic lattice distortion since often such distortion appears in the cuprates. For simplicity, our analysis is carried our for the case when $t'=0$ and the lattice structure is changed by tuning the $t_{0,1}/t_{1,0}$ ratio. One can see that when $t_{0,1}/t_{1,0}\neq 1$, the $d$-$wave$ gap is decreased mainly in the region of SC+N phase stability and the $s$-$wave$ gap component changes sign [see Figs. \ref{fig:dt_dep} (a) and (b)]. In Figs. \ref{fig:dt_dep} (c) and (d) we show how the anisotropy in the hopping integrals affect the anisotropy of the hopping averages in the correlated state $P^G_{0,1}$, $P^G_{1,0}$. As one 
can see, in 
the doping range close to $\delta\approx 0.1$ even for very small lattice anisotropy ($t_{0,1}/t_{1,0}\approx0.95$, $t_{0,1}/t_{1,0}\approx0.97$) we obtain a substantial anisotropy of the hopping averages ($P^G_{0,1}/P^G_{1,0}\approx0.6$). Moreover, as shown before even for the case of square lattice, one obtains $P^G_{0,1}/P^G_{1,0}\lesssim 1$ which signals a spontaneous $C_4$ symmetry breaking [red solid line in \ref{fig:dt_dep} (c)] and leads to the SC+N phase. 
%%%%%%%%%%%%%%%%%%%%%%%%%%%%%%%%FIG4%%%%%%%%%%%%%%%%%%%%%%%%%%%%%%%%%%%%%%%%%%%
%%%%%%%%%%%%%%%%%%%%%%%%%%%%%%%%%%%%%%%%%%%%%%%%%%%%%%%%%%%%%%%%%%%%%%%%%%%%%%%

It is not clear what determines the optimal values of doping which lead to the tendency towards anisotropic character of the electronic properties. Nevertheless, significance of the electronic correlations taken into account by higher order terms of the diagrammatic expansion (\ref{eq:expansion}) is evident, since the analyzed result can be obtained only by going beyond the RMFT approach. We show this in Fig. \ref{fig:dt_dep} (d), where the comparison of the two methods is provided. Since within the RMFT approach no stability of the SC phase is obtained in the Hubbard model, we compare the two methods limiting to the pure nematic phase only. As one can see, for the case of square lattice ($t_{0,1}/t_{1,0}=1$), no nematic 
behavior ($P^G_{0,1}/P^G_{1,0}=1$) is obtained according to the RMFT method, whereas within the DE-GWF approach the anisotropic behavior of the electronic system is sustained. Also, as shown in the inset to Fig. \ref{fig:dt_dep} (d), in RMFT we obtain $P^G_{0,1}/P^G_{1,0}\approx t_{0,1}/t_{1,0}$ in the whole doping range, while the DE-GWF approach leads to a large enhancement of the electronic anisotropy for $\delta\lesssim 0.15$.

%%%%%%%%%%%%%%%%%%%%%%%%%%%%%%%%FIG5%%%%%%%%%%%%%%%%%%%%%%%%%%%%%%%%%%%%%%%%%%%
%%%%%%%%%%%%%%%%%%%%%%%%%%%%%%%%%%%%%%%%%%%%%%%%%%%%%%%%%%%%%%%%%%%%%%%%%%%%%%%

\section{Conclusions}
This paper is a continuation of our detailed studies of high-T$_C$ SC within an extended $t$-$J$ (or extended Hubbard) model treated within the diagrammatic expansion of the Gutzwiller wave function (DE-GWF) in two dimensions, that goes beyond the renormalized mean-field theory in a systematic manner\cite{Spalek2017,Zegrodnik2017_1,Zegrodnik2017_2,Abram2017}. Explicitly, we have analyzed the effect of all the significant intersite interaction terms on the coexistence of superconducting (SC) and nematic (N) phases within that method. As a starting point of our analysis we have determined the stability range of the coexistent phase on the $(\delta,U)$ plane for the case of Hubbard model. The coexistent SC+N phase appears for high enough values of the Coulomb repulsion ($U\gtrsim 6$) and in a wide doping range. Due to the $C_4$ symmetry breaking the $d$-$wave$ pairing amplitude is suppressed and the $s$-$wave$ component of the SC gap appears in the SC+N phase (cf. Fig. \ref{fig:diag_nU}). This signals a competing character of 
the SC and N phases interplay. Moreover, the appearance of the $s$-$wave$ SC component with the onset of nematicity in addition to the $d$-$wave$ SC hampers the gapless character of the latter in the nodal direction. Nevertheless, the $s$-$wave$ amplitude is about two orders of magnitude smaller than that of the $d$-$wave$.  

For the case of the extended model the competition between SC and N is determined by both the exchange interaction and the intersite Coulomb repulsion terms. Namely, the $J$-term enhances SC and suppresses nematicity, whereas for the $V$-term the opposite is true (cf. Figs \ref{fig:J_dep} and \ref{fig:V_dep} (a), (c), (d)). According to our analysis of the $t$-$J$-$U$ model, the nematicity survives up to $J\approx 0.15$ what means that the SC+N phase is already destroyed for the parameter set, for which a good agreement between theory and experiment has been achieved for the copper-based superconductors\cite{Spalek2017}. Hence, in such as situation the $s$-$wave$ gap component is absent (only pure $d$-$wave$ SC survives) and the nodal direction is well defined. Nevertheless, by adding the $V$-term to the $t$-$J$-$U$ model, one could still sustain the stability of the SC+N phase for the values of $J\approx 0.3$, typical for the cuprates. 

Our analysis of the effect of electronic structure details on the SC+N phase have shown that there is no influence of the van Hove singularity position on the lower critical doping for the coexistent-phase onset. For all the considered $t'$ values the lower critical doping remains almost constant and equal to $\delta_c\approx 0.05$, the value close to that, below which the antiferromagnetic phase appears in the cuprates. This result differs with that presented in Ref. \onlinecite{Kitatani2017}, where the FLEX+DMFT method has been used. However, as mentioned earlier, the results obtained within the latter method are limited to small Hubbard-model $U$ values.

The influence of the correlated hopping term on the SC+N phase is not significant up to the value $K\approx 1$, where the nematicity is destroyed and the $d$-$wave$ gap is suppressed (cf. Fig. \ref{fig:K_tp_dep}). 

As could be expected, the assumed from the start anisotropy of the lattice induces anisotropy of the electronic properties in the whole doping range. However, a substantial increase of the electronic anisotropy is obtained close to $\delta\approx 0.1$, both for the case of the coexistent SC+N phase [cf. Fig. \ref{fig:dt_dep} (c)] and for the pure nematic phase [cf. inset to Fig. \ref{fig:dt_dep} (d)]. Such a result brings into mind the experimental data for the cuprates, where a very small structural anisotropy leads to a large effect for selected physical properties\cite{Ando2002,Daou2010,Hinkov2008,Lawler2010}. The latter result is not reproduced within RMFT method, where we obtain $P^G_{0,1}/P^G_{1,0}\approx t_{0,1}/t_{1,0}$ in the whole doping range. Moreover, within the RMFT, no spontaneous $C_4$ symmetry breaking appears for the case of square lattice [cf. Fig. \ref{fig:dt_dep} (d)]. This in turn demonstrates that the correlation effects taken, into account in the higher-order of the DE-GWF approach, are 
responsible 
for the nematic phase appearance for the square-lattice case.

It would be interesting to investigate if the susceptibility towards the $C_4$-symmetry breaking of the electronic system can also induce the orthorhombic crystal distortion of the lattice within the present approach. 
In order to take into account the subtle interplay between the electronic system and the lattice structure, one would have to calculate the hopping integrals in an ab-initio fashion instead of treating them as model parameters as here. Such an analysis could be carried out by combining the DE-GWF method with the EDABI\cite{Biborski2015,Biborski2015_2,Biborski2017} approach. Moreover, such a method could also be used to analyze theoretically the interplay between the unconventional superconductivity and lattice distortion, which is observed in the copper based compounds\cite{Horn1987}.

The question of connection between the nematicity and the CDW appearance is quite involved and requires a separate detailed analysis.

\section{Acknowledgment}

MZ acknowledges the financial support from the Ministry of Science and Higher Education. 
JS acknowledges the financial support through the Grant MAESTRO, No. DEC-2012/04/A/ST3/00342 from the National Science Centre (NCN) of Poland.

\end{document}